\documentstyle[preprint,eqsecnum,aps]{revtex}

\newcommand{\be}{\begin{equation}}
\newcommand{\ee}{\end{equation}}
\newcommand\beq{\begin{eqnarray}}
\newcommand\eeq{\end{eqnarray}}

\begin{document}
\title{Nucleon Solution of the Faddeev Equation\\
       in the Nambu-Jona-Lasinio Model}
\author{Suzhou Huang}
\address{Department of Physics, FM-15\\
         University of Washington\\
         Seattle, WA 98195}
\author{John Tjon}
\address{Institute for Theoretical Physics\\
         Unversity of Utrecht\\
         3508 TA Utrecht, The Netherlands}
\date{\today}
\maketitle

\begin{abstract}
Given the phenomenological success of the Nambu-Jona-Lasinio model in
describing the meson physics in the low energy limit, it is tempting
to find the fully relativistically structured nucleon solution in the
same model under the similar approximation employed in the mesonic
sector. To achieve this goal we need to solve a relativistic Faddeev
equation. The factorizability of the two-body
T-matrix reduces the three-body Faddeev equation to a tractable
two-body Bethe-Salpeter equation. The reduced equation is then
solved numerically. Our result indicates that the nucleon consists of
three loosely bound constituent quarks.
\end{abstract}
\pacs{ }

\widetext

\section{Introduction}

  One of the most important feature of QCD is the chiral symmetry
and its dynamical breaking, which is expected to dictate the low
energy hadronic physics. There exist work,
such as the QCD sum rule \cite{SUMRULE}, the instanton
liquid model \cite{SHURYAK} and an explicit lattice QCD simulation
via cooling technique \cite{COOLING}, directly or indirectly
confirming this expectation. The Lagrangian introduced by Nambu
and Jona-Lasinio \cite{NAMBU} long time ago conveniently mimics
such an essential aspect of QCD in the low energy limit. Models
based on the NJL type of Lagrangians
have been demonstrated to be very successful in describing the
low energy mesonic physics \cite{WEISE}. On the other hand, due
to technical reasons, these models are much less effective in
describing low energy physics involving baryons. It is very often
that extra assumptions beyond these models have to be used
in order to make concrete predictions in the baryonic sector.

  While there is very little doubt that the NJL type of models
could support bound baryonic states, the direct approach in
solving a three-body problem has only been attempted recently
\cite{BUCK} with approximations apparently quite different from
that of employed in the mesonic sector.
An important point is that approximations
in the baryonic sector have to be consistent with chiral symmetry,
for example, the nucleon solution should approximately satisfy
the Goldberger-Treiman relation \cite{GT}. Otherwise the very
essence of the NJL model, the chiral symmetry, is ruined by the
ad hoc approximations. Other indirect attempts in finding the
nucleon solution in the NJL-like models, such as the non-topological
soliton approach \cite{NON-T}, the bosonization approach
\cite{BOSON} and undoubtedly others can be found in the literature.

  In this paper we undertake the task of finding
a nucleon-like solution in the NJL type of models. First we
derive the three-body Faddeev equation in the valence constituent
quark approximation by ignoring the three-body irreducible graphs.
Due to the heavyness of the constituent quark, this approximation
is expected to be good at low energies, as shown in the mesonic
sector. By observing that the two-body diquark $T$-matrix has a
separable form the Faddeev equation can be reduced to an effective
two-body Bethe-Salpeter equation with an energy dependent
interaction. Then the reduced problem
is solved numerically, without any further approximations.
Although we can not explicitly show that our solution respects
the exact chiral symmetry, in contrast with the meson solutions
in the Hartree-Fock approximation, we believe that our work is
a step forward in the right direction. So long as we can find
weakly bound nucleon-like state of three constituent quarks, the
chiral symmetry should be well protected, since the chiral symmetry
is exact at the constituent quark level \cite{WEISE}.

  This paper is organized as follows. In section 2 we first introduce
the model we explicit consider and then briefly review the two-body
sector to fix parameters in the model. In section 3 the derivation
of the three-body Faddeev equation and its reduction to the effective
two-body equation are presented. The numerical technique involved
in solving the reduced fully relativistic Bethe-Salpeter equation,
based on the work of Rupp and Tjon \cite{TJON}, is recapitulated and then
applied to our case in section 4. A summary and some outlook follows
in section 5.

\section{Two-body Sector}

The Lagrangian we consider is the two flavored Nambu-Jona-Lasinio
model given by
\begin{equation}
{\cal L}=\bar{\psi}i\gamma_\mu\partial^\mu\psi
+G_1[(\bar{\psi}\psi)^2+(\bar{\psi}i\gamma_5\tau_a\psi)^2]
-G_2[\bar{\psi}\gamma_\mu(\lambda_A/2)\psi]^2,\label{LAG}
\end{equation}
where $\psi$ is the quark field, $\tau_a$ ($a=1,2,3$) and
$\lambda_A$ ($A=1,2,\cdots,8$) are the generators of the
flavor $SU_f(2)$ and color $SU_c(3)$ groups respectively.
Small current quark masses are ignored for simplicity. Since
the coupling constants $G_1$ and $G_2$ have negative mass
dimension, this model is not renormalizable. An appropriate
ultraviolet cutoff procedure has to be specified in order to
make the model well defined. In this work we insert a form
factor $g(k)=g(-k)$, whose functional form will be eventually
taken to be a four-momentum cutoff $\Lambda$ in Euclidean space
for convenience, at every fermion vertex in the loop integrals.

The justification of using Eq.(\ref{LAG}) to model the low energy
physics of the strong interaction and its phenomenological success
in mesonic channels were well studied in the literature. A recent
review can be found in reference \cite{WEISE}. Although our
primary goal is to find three-body baryonic solutions in this
model, it is adequate to recapitulate the essential features of
this model in the meson sector, which is used to fix all the
parameters but one in the model. Then the two-body $T$-matrix in
the scalar-isoscalar diquark channel, which consists of an essential
component of the three-body Faddeev equation, will be derived.

\subsection{Meson Channel}

The most important feature that makes the model resembles QCD
at low energy domain is that the NJL model and QCD share the
same chiral symmetry and its dynamical breaking. The manifestation
of this phenomenon in the NJL model is that the massless quarks
acquire dynamical masses through the following self-consistent
gap equation, when only the fermion bubble chain graphs are
included or in the Hartree-Fock approximation,
\begin{equation}
1=i(a_1G_1+a_2G_2)\int {d^4k\over (2\pi)^4}
{4g(k)\over k^2-m^2},\label{GAP}
\end{equation}
where $m$ is the constituent quark mass, which is related
to the fermion condensate $\langle\bar{\psi}\psi\rangle$ by
$m=-(a_1G_1+a_2G_2)\langle\bar{\psi}\psi\rangle/(N_cN_f)$.
For $N_c=3$ and $N_f=2$, $a_1=13$ and $a_2=8/3$.

As a consequence of the chiral symmetry breaking the pion emerges
as the massless Goldstone boson, which manifests itself explicitly as the
massless pole in the quark-antiquark two-body $T_{\bar{q}q}$-matrix
in pseudoscalar channel. If again only the fermion bubble chain
graphs are retained or in the RPA approximation, $T_{\bar{q}q}$
given by Fig.~\ref{F:PION} can be readily calculated. The residue
of $T_{\bar{q}q}$-matrix at this pole, $\Gamma_\pi^a$, has the form
\begin{equation}
\Gamma_\pi^a=g_{\pi \bar{q}q}[\openone_C\otimes\tau^a\otimes i\gamma_5],
\end{equation}
where $g_{\pi\bar{q}q}$ is the pion-quark-antiquark coupling constant.
The pion decay constant, $f_\pi$, is defined through the axial-vector
current matrix element,
\begin{equation}
i f_\pi p_\mu \delta_{ab}\equiv
\langle 0 |\bar{\psi}\gamma_\mu\gamma_5 {\tau_a\over 2}\psi
|\pi_b(p)\rangle.
\end{equation}
Using the chiral Ward identity, or the Goldberger-Treiman relation
at the quark level, $f_\pi g_{\pi\bar{q}q}=m$, one can easily find
\begin{equation}
f_\pi^2=4N_c m^2 \int{d^4k\over(2\pi)^4}
{ig(k)\over [ k^2-m^2 ]^2}. \label{FPION}
\end{equation}
In arriving at the above result the on-shell condition $p^2=m_\pi^2=0$
has been used.

There are three parameters in the model,
two couplings $G_1$, $G_2$ and the cutoff $\Lambda$. By equating $f_\pi$
and $m$ or $\langle\bar{\psi}\psi\rangle$ to the phenomenological
values through Eq.(\ref{GAP}) and Eq.(\ref{FPION}) we can fix two
of them, which we pick $\Lambda$ and $G\equiv a_1 G_1+a_2 G_2$.
This more or less fixes the theory in the mesonic sector. The last
parameter $\eta\equiv G_1/G_2$ is left free to vary.

\subsection{Diquark Channel}

If we use the same fermion bubble chain approximation in the
quark-quark sector, we can easily calculate the corresponding
$T$-matrix. In the color $\bar{3}$ scalar-isoscalar channel
the $T$-matrix has the structure, when ignoring the mixing
with other channels (for example, the color $\bar{3}$
vector-isoscalar channel)
\begin{equation}
T_{qq}^{ab,dc}(p)=
i R(p)[\lambda_A^{[\,\,]}\otimes\tau_2\otimes C\gamma_5]^{ab}
[\lambda_A^{[\,\,]}\otimes\tau_2\otimes C\gamma_5]^{dc},
\label{DIQUARK}
\end{equation}
where $\lambda_A^{[ij]}\equiv (\lambda_A^{ij}-\lambda_A^{ji})/2$,
$C\equiv i\gamma_0\gamma_2$ is the charge conjugation matrix and
$a,b$ label all the color, flavor and Dirac indices. The scalar
function $R(p)$ can be obtained straightforwardly by summing the
fermion bubble chain, yielding $R(p)=G'/(1-G' J(p))$ with
$G'=(b_1G_1+b_2G_2)/4$ and
\[
J(p)=4i\int {d^4k\over(2\pi)^4} g^2(k)\,
{\rm Tr}[C\gamma_5 S_F(k+p/2) C\gamma_5 S_F^T(-k+p/2)],
\]
where ${\rm Tr}$ denotes the trace in Dirac space and
$S_F(k)$ and $S_F^T(k)$ are the constituent quark propagator
and its transpose (in Dirac space) respectively. Furthermore we
have $b_1=4$, $b_2=8/3$.

Whether there exists diquark bound states in this model depends on
whether $R(p)$ develops poles in the time-like region. As shown in
\cite{ZAHED} it is possible by varying $\eta$ to find a bound
diquark state in this channel. It should be emphasized
that the existence of such a diquark bound state is not a necessary
condition for the existence of a three-quark bound state, though
it might be useful to utilize the diquark concept phenomenologically
to explain certain scaling violations in lepton-nucleon experiments.
In this paper the diquark state is merely an intermediate device in
setting up the three-body Faddeev equation. The phenomenological
relevance of the diquark will not be pursued here.

\section{Three-body Sector}

Given the fundamental four-fermion vertex by the Lagrangian and
the quark-quark two-body $T_{qq}$-matrix, and ignoring the three-body
irreducible graphs, the three-body $T$-matrix can be solved from
the Faddeev equation by iterating the fundamental vertex and the
two-body $T_{qq}$-matrix. Throwing away the three-body irreducible
graphs is in some sense equivalent to ignoring the non-valence
constituent quark loops in the iteration process. Due to the heavyness
of the constituent quark mass ($300\sim400$MeV) this approximation is
justified in the low energy region. Of course, one should realize that
we do not invoke more approximation here. Essentially the same kind of
approximation was used in the mesonic and diquark cases.

\subsection{Faddeev Equation}
Since we are only interested at the moment in the three-body bound
state, we only need to consider the homogeneous Faddeev equation.
If the full three-body amplitude $\Gamma^{f,d}$ (with $f$ and $d$
being the external flavor and Dirac indices) is decomposed as a
sum of three partial amplitudes $\Gamma_i$ ($i=1,2,3$), with
\begin{equation}
\Gamma_1^{f,d}\equiv\epsilon_{c_1c_2c_3}\tau_2^{f_2f_3}
(C\gamma_5)^{d_2d_3}\delta^{f_1f}
{\Gamma}_{d_1d}^{(1)}(p_1,p_2,p_3),
\end{equation}
and similarly for $\Gamma_2$ and $\Gamma_3$ by cyclically permuting
$(1,2,3)$, then these partial amplitudes satisfy the following
integral equation,
\begin{eqnarray}
\lefteqn{{\Gamma}^{(3)}(p_1,p_2,p_3)
=2g(p_1-p_2) iR(p_1+p_2)\times } \nonumber \\
&&\left\{ \int {d^4p_1'\over(2\pi)^4} g(p_1'-p_2')
   [C\gamma_5 S^T_F(p_2')C\gamma_5 S_F(p_1')]
   {\Gamma}^{(1)}(p_1',p_2',p_3) \right.
\label{FADDEEV} \\
&&+\left. \int {d^4p_2'\over(2\pi)^4} g(p_1'-p_2')
   [C\gamma_5 S^T_F(p_1')C\gamma_5 S_F(p_2')]
   {\Gamma}^{(2)}(p_1',p_2',p_3) \right\}. \nonumber
\end{eqnarray}
The notation of the above equation is depicted in Fig.~\ref{F:FADDEEV}.
The factor of 2 in Eq.(\ref{FADDEEV}) arises from the color sum
$\epsilon_{ac_1c_2}\epsilon_{bc_1c_2}=2\delta_{ab}$.
Though formally similar to the non-relativistic Faddeev equation,
Eq.(\ref{FADDEEV}) is exact within the approximation mentioned
above. An analogous equation with scalar particles was considered
by Rupp and Tjon in a different context \cite{TJON}. Since we
explicitly included the color, flavor and Dirac structures in
the definition of the two-body $T_{qq}$-matrix and three-body
amplitudes $\Gamma_{1,2,3}^{f,d}$, the recoupling-coefficient
matrix has already been automatically taken into account in
Eq.(\ref{FADDEEV}).

\subsection{Reduction to an effective Bethe-Salpeter equation}

If the two-body $T_{qq}$-matrix involved has a general form, it would
be a formidable task to find the solution for Eq.(\ref{FADDEEV}).
The crucial observation is that $T_{qq}$-matrix has a factorized form
and hence we are only dealing with the so-called separable situation.
The separatability of the two-body interaction leads to a reduction
of the three-body problem to an effective Bethe-Salpeter equation.
As a matter of fact, this reduction has already been hinted by the
explicit form of Eq.(\ref{FADDEEV}). More concretely, the three-body
amplitudes can be written as
\begin{equation}
{\Gamma}^{(1)}_{dd'}(p_1,p_2,p_3)=\Psi_{dd'}(p_1)
g(p_2-p_3)R(p_2+p_3),
\end{equation}
and similarly for ${\Gamma}^{(2,3)}$, with $\Psi$ satisfying
\begin{equation}
\Psi(p_3)=4i\int{d^4p_1'\over(2\pi)^4}
g(p_1'-p_2')R(p_2'+p_3)g(p_2'-p_3)
[C\gamma_5S_F^T(p_2')C\gamma_5 S_F(p_1')]\Psi(p_1'),\label{EBS}
\end{equation}
as a matrix equation in Dirac space. When deriving the above
equation the quarks are treated as identical particles.

Diagrammatically, Eq.(\ref{EBS}) can be represented by
Fig.~\ref{F:EBS}, which looks like a boson (with propagator $R$)
coupling to a third quark to form a three-body bound state.
However, this ought to be distinguished from identifying the
``boson'' as the diquark bound state. The reduction of the
three-body Eq.(\ref{FADDEEV}) to the effective two-body
Eq.(\ref{EBS}) does not depend on whether the diquark channel
has a pole, but rather on the separatability of $T_{qq}$-matrix.

Introducing the equal-mass Jacobi momentum variables $q$ and
$q'$,
\begin{equation}
\label{jac1}
p_3\equiv {P\over 3}-q;\,\,\,\,\,
p'_1\equiv {P\over 3}-q';
\end{equation}
where the total momentum $P$ is given by
\begin{equation}
\label{jac2}
P\equiv p_1+p_2+p_3=p_1'+p_2'+p_3;
\end{equation}
we find that the reduced Bethe-Salpeter equation can be written as
\be
\label{bsf}
\Psi(P,q) = \frac{i}{4 \pi^4} \int d^4q'  V(q,q';P)
R({2\over3} P +q') K \Psi(P,q'),
\ee
where we have defined an energy dependent interaction
\be
\label{int1}
V(q,q';P) =
\frac{g(p_1'-p_2') g(p_2'-p_3)}{(p^{'2}_1-m^2)(p^{'2}_2-m^2)}.
\ee
The Dirac structure of the kernel is contained in the operator
\be
\label{kernel1}
K = [ C \gamma_5 ( \gamma^T p'_2 + m) C \gamma_5  (\gamma p'_1 + m) ].
\ee
Using well-known properties of the charge conjugation operator
$C$, this simplifies to
\be
K =  (   \gamma p'_2 + m ) ( \gamma p'_1 + m).
\ee
In view of Eqs.~(\ref{jac1}-\ref{jac2}) the various momenta present in
Eq.~(\ref{bsf}) can be expressed in terms of the Jacobi
variables $q$, $q'$ and total momentum $P$.

\subsection{Decomposition of the Reduced Amplitudes}

To see the Dirac structure more clearly let us reduce
the operator K into the Pauli form. Using the
$\rho$ -spin notation of Ref.~\cite{gammel} for the upper and
lower compoenents of four-spinors, we get for the matrix
elements $K(\rho,\rho')$ (with $\rho,\rho'=\pm$)
\beq
\label{pauli}
K(+,+) &=& (p'_{20}+m)(p'_{10}+m) - \vec{\sigma}.\vec{p}_2'
\vec{\sigma}.\vec{p}_1' \\
\nonumber
K(+,-) &=& -(p'_{20}+m) \vec{\sigma} . \vec{p}_1' -(m-p'_{10}) \vec{\sigma} .
\vec{p}~'_2 \\
\nonumber
K(-,+) &=& (p'_{10}+m) \vec{\sigma} . \vec{p}_2' -(m-p'_{20}) \vec{\sigma} .
\vec{p}_1' \\
\nonumber
K(-,-) &=& (-p'_{20}+m)(-p'_{10}+m) - \vec{\sigma}.\vec{p}_2'
\vec{\sigma}.\vec{p}_1'
\eeq
The simplest approximation which can be made is to neglect the
lower components, i.e. the kernel is replaced by  $K(+,+)$.
The resulting eigenvalue equation becomes in this case
\be
\label{1ch}
\frac{i}{4 \pi^4} \int d^4 q' V(q,q';P) R(\frac{2}{3} P +q')
[ ( p'_{20}+m)(p'_{10}+m) - \vec{\sigma}.\vec{p}~'_2
\vec{\sigma}.\vec{p}~'_1 ]
\chi(q') = \lambda \chi(q),
\ee
where a physical bound state solution corresponds to the eigenvalue
$\lambda=1$.
Assuming we are in the overall three-quark c.m. system $P=
(\sqrt{s},{\vec 0})$, we see that there are two classes of
solutions to Eq.~(\ref{1ch})
\beq
\label{phi1}
\chi_1 &=& \Phi_1(q_0,|\vec{q}|) \\
\nonumber
\chi_2 &=& \vec{\sigma} . \vec{q}~~ \Phi_2 (q_0,|\vec{q}|)
\eeq
These classes are not coupled to each other in the integral
equations. This is due to parity and angular momentum
conservation. $\chi_1$ is a s-wave solution with $(l=0, s=1/2,
j=1/2)$ and $\chi_2$ is a p-wave $(l=1,s=1/2,j=1/2)$.
In view of the simple form of the $\chi'$s the angular
integration in Eq.~(\ref{1ch}) can explicitly be carried out.
As a result, we obtain a two-dimensional integral equation of the form
\begin{equation}
\chi_n(q)=\frac{i}{ 2  \pi^3} \int_{-\infty}^{\infty} dq'_0
\int_{0}^{\infty} q'^2 dq' V_n(q,q';P) R({2 \over 3} P+q')
\chi_n(q')
\end{equation}
where $q\equiv |{\vec q}|$ and
\begin{equation}
V_n(q,q';P)=  \int_{-1}^{1} dx
\frac{g(p_1'-p_2') g(p_2'-p_3)}{(p^{'2}_2-m^2)(p^{'2}_1-m^2)}
{\rm Tr}_2 [ O_n K(+,+)]
\end{equation}
with $x=cos(\theta_{{\bf q}{\bf q'}})$ and
${\rm Tr}_2$ is the trace to be taken in Pauli space.
Furthermore, the operator $O_n$ is given by $1/2$ and
$(\vec{\sigma}.\vec{q'})/(2 \vec{q}^2)$ for n=1 and 2 respectively.

This analysis can be extended to the full equation.  From
Eq.~(\ref{pauli}) we see that the Pauli spin dependence in $\Psi$ can be
either the unit operator or ${\vec \sigma}.{\vec q}$.
In view of parity conservation there are also two
classes of solutions, which are given by four spinors of the
form
\be
\label{psi1}
\Psi_1 = \left( \begin{array}{c} \phi_1(q_0,q) \\
        \vec{\sigma} . \vec{q}~~ \phi_2(q_0,q) \end{array} \right )
\ee
and
\be
\Psi_2 = \left( \begin{array}{c} \vec{\sigma} . \vec{q}~~ \phi_3(q_0,q)
\\ \phi_4(q_0,q) \end{array} \right )
\ee
The vertex functions $\Psi_1$ and $\Psi_2$ are again not coupled
to each other.  With this form  for $\Psi$ a
partial wave decomposed set of coupled integral equations can be
derived. Inserting Eq.~(\ref{psi1})  in Eq.~(\ref{bsf}) we get
\begin{equation}
\label{2ch}
\phi_n(q) = \frac{i}{ 2 \pi^3}
\sum_{m=1}^2 \int_{-\infty}^{\infty} dq'_0
\int_{0}^{\infty} q'^2 dq' V_{nm}(q,q';P) R(\frac{2}{3} P +q')
 \phi_m(q')
\end{equation}
with $n=1,2$ and
\begin{equation}
V_{nm}(q,q';P)= \int_{-1}^{1} dx
\frac{g(p_1'-p_2') g(p_2'-p_3)}{(p^{'2}_2-m^2)(p^{'2}_1-m^2)}
K_{nm}
\end{equation}
The explicit expression for the matrix $K_{nm}$
can be determined by
noting that
\be
\Psi_1 = [\frac{1+\gamma_0}{2} \phi_1  - \vec{\gamma} . \vec{q}
\frac{1+\gamma_0}{2} \phi_2 ] w,
\ee
where $w$ is a four-spinor with every component equal 1.
The operator $(1+\gamma_0)/2$ is
in the three-quark c.m. system nothing else as the projection
operator $\Omega=(M_N+\gamma P)/(2 M_N)$.
With this we can now calculate $K_{nm}$ by projecting out the
Dirac form  on $\Omega$ and $\vec{\gamma}.\vec{q}$. In so doing
we get for the Dirac part of the kernel
\beq
\label{k2}
K_{1m} &=& \frac{1}{2} {\rm Tr}[ K \kappa_m(\vec{q}~')]
\\
\nonumber
K_{2m} &=& \frac{{\rm Tr}[ \vec{\gamma}.\vec{q} K \kappa_m(\vec{q}~')]}
{{\rm Tr}[\vec{\gamma}.\vec{q}\kappa_2(\vec{q})]}
\eeq
where $\kappa_1=\Omega, \kappa_2=\vec{\gamma}. \vec{q} \,\Omega$ .
Eq.~(\ref{k2}) can be evaluated in a straightforward way. We
find

\beq
 K_{11} &=& - q_0 q_0~' + q_0   ( m + \frac{M_N}{3} )
   - q~'^2-q.q~' + (m + \frac{M_N}{3})^2
\\
\nonumber
   K_{12} &=& - ( q_0   - 2 \frac{M_N}{3} )   {\vec q}~'^2
   + ( q_0~' + m - \frac{M_N}{3} )   ({\vec q} . {\vec q}~')
\\
\nonumber
   K_{21} &=&  q_0~'  - m - \frac{M_N}{3} -
(q_0 + 2 \frac{M_N}{3}) \frac{{\vec q} . {\vec q}~'}{{\vec q}^2}
\\
\nonumber
   K_{22} &=&  {\vec q}~'^2 +[- q_0 (q_0~' - m + \frac{M_N}{3})
     - q'^2  + (m - \frac{M_N}{3})^2 ]
\frac{{\vec q} . {\vec q'}}{{\vec q}~^2}
\eeq
In a similar way the coupled set of equations can be derived for
$\Psi_2$. It should be noted that possible solutions of this type
correspond to p-wave like states and as a result are expected
not to be the ground state of the three quark system
due to the centrifugal term.
Since we are interested in this paper in the
nucleon, it is natural to confine ourself to the solutions of
the s-wave type, given by $\Psi_1$.

\section{Calculations}

Following Ref.~\cite{TJON} the resulting integral equations can
be studied by performing a Wick rotation of the $q_0$ and $q_0'$
variables to the complex plane. Assuming that the diquark system
supports a boundstate at $M_{qq}$,
we find that at the threshold point of quark-diquark scattering a
pinching singularity can occur in the kernel of Eq.~(\ref{2ch})
at $q_0=\hat{q}_0=\frac{1}{3}(2m-M_{qq})$. It can readily be
verified that in the triquark boundstate region, corrresponding
to $\sqrt{s}< m_q+M_{qq}$,
the $q_0$ and $q_0'$ variables can be rotated
to a path going through the point $\hat{q}_0$ and
parallel to the imaginary axis without encountering any
singularities in the kernel.
In so doing we implicitly assume that eventual
singularities in the form factors $g(q)$ do not cross the
imaginary $q_0$ axis. Furthermore the arguments of the form
factors are approximated by neglecting the $\hat{q}_0$
dependence. The resulting Euclidean form of the
integral equation is regular in the boundstate region and as a
result it can in principle be solved by standard discretization
procedures. Because of the actual size of the resulting matrix
equations we have adopted the method described in
Ref.~\cite{TJON}. The perturbation series is determined by
iterating the equations, while the occurring
two-dimensional integrals are evaluated using standard Gaussian
quadratures. From this series the energy
of the boundstate is determined using the ratio method of
Malfliet and Tjon~\cite{malf}. It should noted that as a byproduct also
the corresponding wavefunction can be found in this way.

There are several parameters in the model: the cutoff $\Lambda$,
the coupling constants $G_1$ and $G_2$.
The overall mass scale can be set by the choice of the
the cutoff mass $\Lambda$. There are two constraints we would
like to satisfy. From the pion decay constant $f_\pi=93 MeV$,
we can determine according to Eq.~(\ref{FPION})
the constituent quark mass $m$.
Taking a value of $\Lambda=750MeV$ we find $m=375MeV$.
Decreasing $\Lambda$ for instance to 739 MeV, the quark
mass increases to $m=400$ MeV.
Secondly, from the self-consistent mass gap equation, the value of
$G \equiv 13 G_1 + 8/3 G_2$ is fixed.
As a consequence the only free parameter is the ratio $G_1/G_2$,
which can be used as the parameter to determine the diquark mass.
In Fig.~\ref{F:diquark} the diquark mass dependence on
this ratio is shown.

Once the parameters of the model have been fixed we may
study the three quark bound state.
In Table \ref{tab1} we list the diquark masses needed to get
a nucleon solution at $M_N=939$ MeV in three
approximations, non-relativistic (or static limit)
$K_{11}\rightarrow 4m^2$; one channel defined by Eq.(\ref{1ch})
and two-channel defined by Eq.(\ref{2ch}). As one can see,
a stable nucleon solution always requires the scalar-isoscalar
diquark state lie below two-quark threshold.
The binding of the three quark system clearly depends
on the choice of the diquark energy. In Fig.~\ref{F:nucleon}
are shown for two cases of $\Lambda$ the results of the
calculated mass of the three quark ground state
as a function of the diquark mass (solid line).
Also are plotted  the results when we only keep the s-wave
components of the three quark wave function (dot-dash line)
and the static limit of $K_{11}\rightarrow 4 m^2$ (dash line).
{}From this we see that at lower diquark masses the static limit
predicts a substantially deeper binding than the full
2-channel result and hence it can be an unreliable approximation.

  To have a feeling on the quality of the static and one channel
approximations we list in Table \ref{tab2} the nucleon masses
with diquark mass fixed at the value where the two-channel
calculation would yield $M_N=939$ MeV. It is clear from Table
\ref{tab2} that the relativistic forms give rise to less attraction,
leading to a slightly higher lying ground state, though there is no
qualitative difference from the static approximation \cite{BUCK}.
{}From these results we may conclude that
over a wide range of $M_{qq}$ a stable nucleon solution
indeed exists in the considered NJL model.
Increasing the diquark mass leads to a weakening of the
quark-quark interaction and as result the nucleon mass increases.

  In the range of diquark mass we considered, the existence of
the nucleon solution near its experimental value required about
150 to 300 MeV binding in the scalar-isoscalar diquark.
This kind of diquark
clustering is also observed in a recent instanton model calculation
by Shuryak et al \cite{instanton} in the nucleon channel and
qualitatively confirmed by the lattice simulation through cooling
\cite{cooling1}. Since the NJL type of models are practically
effective theories for these instanton models, the similar diquark
clustering in our case may not be a mere coincidence.

\section{Summary and Outlook}

  We have been able to demonstrate that the NJL type of models
can easily accommodate the nucleon-like state under similar type
approximations employed in the mesonic sector. Although we were
not able to explicitly show, from the derived Faddeev equation,
that the solution of the nucleon state satisfies the
Goldberger-Treiman relation, we indeed found that the nucleon
be a loosely bound state of the constituent quarks. In order to
have the nucleon solution as a true bound state a bound diquark
in the scalar-isoscalar channel is necessary in our model.

There are clearly some interesting questions which can be addressed
in such a model.  Using the wavefunction corresponding to the
three-quark boundstate, the properties of the various form factors
for the nucleon can in principle be studied. It is also of interest
to examine possible Delta isobar states in the same model we have
considered. The mass splitting between the baryon decuplet and
octet constitutes a non-trivial test of the NJL type of models,
while the mass splittings within the same baryon multiplets are less
stringent due to the fact that the latter splittings mainly come from
quark masses. Since the dominant diquark configuration in the Delta
should be vector-isovector, the resulting three-body bound
state could be a resonance rather than a bound state. The numerical
method we used in the nucleon case needs to be modified if the
Delta lies in the continuum. A more careful examination of the
compatibility of the Faddeev equation and the chiral symmetry
could provide useful insight on how the Goldberger-Treiman relation
at the nucleon level is realized. Finally, also
pion-nucleon and nucleon-nucleon low energy scattering processes can in
principle be studied.
It is easy to anticipate that the meson-exchange potential could merge
between nucleons, if only the valence quark lines are included at each
instance in the Feynman graphs. Furthermore, since the nucleon in
this model is a loosely bound state of three constituent quarks,
it is very likely that there are anomalous singularities in the
scattering processes, which could potentially modify the one-meson
exchange nuclear force picture in the low energy limit. The role
of the anomalous singularity in the form factor for loosely
bound states in similar models was studied recently \cite{HL}.

{\bf Acknowledgements}

This work is supported in part by funds
provided by the U. S. Department of Energy.

\begin{figure}
\caption{Feynman graphs for $T_{\bar{q}q}$ in pseudoscalar channel.}
\label{F:PION}
\end{figure}

\begin{figure}
\caption{Feynman graphs for $T_{qq}$ in scalar-isoscalar diquark channel.}
\label{F:DIQUARK}
\end{figure}

\begin{figure}
\caption{Faddeev equation for the nucleon.}
\label{F:FADDEEV}
\end{figure}

\begin{figure}
\caption{Reduced effective Bethe-Salpeter equation for the nucleon.}
\label{F:EBS}
\end{figure}

\begin{figure}
\caption{Scalar-isoscalar diquark mass as a function of $G_1/G_2$
for two values of constituent quark masses. The horizontal lines
indicate the quark-quark thresholds.}
\label{F:diquark}
\end{figure}

\begin{figure}
\caption{Nucleon mass as a function of the diquark mass,
with the dash line corresponding to static limit
$K_{11}\rightarrow 4m^2$, dot-dash line corresponding to
one channel defined by Eq.~(3.12), solid line correspoding to full
two-channel defined by Eq.~(3.18) solutions respectively. The dotted
line indicates the quark-diquark scattering threshold.}
\label{F:nucleon}
\end{figure}

\begin{table}
\caption{Diquark masses needed to get a nucleon solution at
$M_N=939$ MeV in various approximations.}
\begin{tabular}{ccccc}
$m$ (MeV) & $\Lambda$ (MeV) & $M_{qq}$\tablenote{static limit.} (MeV)
& $M_{qq}$\tablenote{one channel defined by Eq.~(\ref{1ch}).} (MeV)
& $M_{qq}$\tablenote{two-channel defined by Eq.~(\ref{2ch}).} (MeV)\\
\tableline
375 & 750 & 579.0 & 570.9 & 576.8 \\
400 & 739 & 572.8 & 554.7 & 564.7 \\
450 & 728 & 577.0 & 531.5 & 547.8 \\
\end{tabular}
\label{tab1}
\end{table}

\begin{table}
\caption{Comparison of predictions of the nucleon mass
in various approximations. The value of $M_{qq}$ is fixed
so that $M_N=939$ MeV in the full two-channel calculation.}
\begin{tabular}{ccccc}
$m$ (MeV) & $\Lambda$ (MeV)
& $M_{qq}$ (MeV)
& $M_N$\tablenote{static limit.} (MeV)
& $M_N$\tablenote{one channel defined by Eq.~(\ref{1ch}).} (MeV)\\
\tableline
375 & 750 & 576.8 & 936.4 & 945.3 \\
400 & 739 & 564.7 & 928.2 & 950.0 \\
450 & 728 & 547.3 & 892.7 & 957.9 \\
\end{tabular}
\label{tab2}
\end{table}

\end{document}